\documentclass[aps,prl,twocolumn,showpacs]{revtex4}
\usepackage{amsfonts}
\usepackage{amsmath}
\usepackage{amssymb}
\usepackage{graphicx}
\usepackage{color}
\usepackage{epsfig}
\usepackage{remreset}
\usepackage{hyperref}
\usepackage[usenames,dvipsnames,svgnames,table]{xcolor}

\hypersetup{colorlinks=true, citecolor=red, linkcolor=blue, urlcolor=violet}

\begin{document}

\title{Relativistic Gravity and Parity-Violating Non-Relativistic Effective
Field Theories}
\author{Chaolun Wu$^{1}$}
\email{chaolunwu@uchicago.edu}
\author{Shao-Feng Wu$^{2,1}$}
\email{sfwu@shu.edu.cn}
\affiliation{$^{1}$Kadanoff Center for Theoretical Physics and Enrico Fermi Institute,
University of Chicago, Chicago, Illinois 60637, USA}
\affiliation{$^{2}$Department of Physics, Shanghai University, Shanghai 200444, China}

\begin{abstract}
We show that the relativistic gravity theory can offer a framework to
formulate the non-relativistic effective field theory in a general
coordinate invariant way. We focus on the parity violating case in 2+1
dimensions which is particularly appropriate for the study on quantum Hall
effects and chiral superfluids. We discuss how the non-relativistic
spacetime structure emerges from relativistic gravity. We present covariant
maps and constraints that relate the field contents in the two theories,
which also serve as the holographic dictionary in context of gauge/gravity
duality. A low energy effective action for fractional quantum Hall states is
constructed, which captures universal geometric properties and generates
non-universal corrections systematically. We give another holographic
example with dyonic black brane background to calculate thermodynamic and
transport properties of strongly coupled non-relativistic fluids in magnetic
field. In particular, by identifying the shift function in the gravity as
minus of guiding center velocity, we obtain the Hall viscosity with its
relation to Landau orbital angular momentum density proportional to Wen-Zee
shift. Our formalism has a good projection to lowest Landau level.
\end{abstract}

\maketitle

\textbf{\emph{Introduction}} Combined with effective field theory (EFT)
techniques, symmetry plays an important role for understanding strongly
correlated systems from high energy to condensed matter and atomic physics.
One preeminent example is the fractional quantum Hall (FQH) effect, where
interactions are crucial and defy perturbative approaches. Based on the
non-relativistic (NR) general coordinate invariance (GCI) introduced in his
seminal paper \cite{Son06}, recently Son studied the system coupled to
Newton-Cartan (NC) geometry \cite{Son13} and constructed an EFT for FQH
states. Of particular interest is that this theory has a good projection to
the lowest Landau level (LLL) and encodes universal geometric properties
such as Hall viscosity \cite{Avron95,Tokatly07,Read:09,Read:11,Hoyos:2011ez}%
. Further developments along this line can be found in \cite%
{Golkar13,Hofmann13,Gromov1403,Banerjee14,Geracie:2014,Brauner:2014jaa,Jensen:2014aia,Andreev:2014gia}%
. It is worthy to note that speaking of EFTs, we also include holography
\cite{Maldacena00}, which can be viewed as EFTs constructed in the aid of
higher dimensional geometries and holographic dictionaries.

For NR field theories, a fundamental feature is the existence of a global
time, a requirement of the NR causality. Thus to build up NR GCI EFTs, it is
natural to employ NR gravity theories such as NC geometry \cite%
{Son13,Geracie:2014,Gromov1403} and Ho\v{r}ava gravity \cite%
{Karch12,Janiszewski:2012nf,Hofmann13,Wu1}. Relativistic gravity theories do
not have a built-in notion of global time \emph{a priori}. However, this
does not exclude the possibility that they can be used as a framework to
construct NR GCI EFTs, provided that the background isometry or certain
imposed condition selects a preferred time foliation. The pioneering works
of \cite{Son08,Balasubramanian08} and \cite{Kachru:2008yh} show that this
can be achieved in relativistic gravity with Schr\"{o}dinger and Lifshitz
backgrounds. Furthermore, \cite{Christensen13} shows that for Lifshitz
holography, the boundary geometry is of various types of NC geometry,
depending on the boundary condition the time-like vielbein satisfies.
Actually such boundary condition may exist independently of the holographic
bulk structure. Our first step in this letter is to argue that relativistic
gravity can be used for general NR EFTs even without the aid of holography,
provided that the time-like vielbein satisfies the hypersurface
orthogonality condition.

A second common feature in the recent geometric formalism of NR EFTs is a
velocity field. In NR field theories, the $U(1)$ gauge field $A_{\mu }$
transforms under diffeomorphism not just as a Lie derivative but with extra
terms only dependent on metric. This is a consequence of Galilean boost
invariance and gives rise to the well-known relation between momentum
density and conserved current \cite{Greiter89} and its variations \cite%
{Geracie:2014}. In EFTs the velocity field is necessary to cancel the extra
terms and covariantize the gauge field. In holography, the relation between $%
A_{\mu }$ and its covariantized version (called as covariant map) serves as
part of the holographic dictionary \cite{HolographicDictionary}. The precise
form of this map may vary depending on how the microscopic theory is coupled
to curved space. In this letter, in addition to local $U(1)$ gauge symmetry
and spatial diffeomorphism widely studied in NR EFTs, we also consider
homogeneous time reparametrization and local anisotropic Weyl rescaling.
These almost completely determine the covariant map. As an example, we show
a single Chern-Simons term together with the covariant map can reproduce all
correlation functions obtained in \cite{Son13} using NC formalism.

A NR EFT formalism built on relativistic gravity is particularly convenient
for applications of holography, which is developed mostly within the frame
of relativistic gravity theories. As a prerequisite for NR holography, a
notion of global time must exist at the boundary. In \cite{Karch12,Wu1} the
global time is extended to the whole bulk by employing Ho\v{r}ava gravity
\cite{Horava09}. However, Ho\v{r}ava gravity is notorious for complications
involving black hole event horizon \cite{Eling06} and difficulty of finding
hyperbolic black hole solutions \cite{Janiszewski:BH}. Moreover, it is not
clear that whether Ho\v{r}ava gravity accommodates the dynamical exponent $%
z\rightarrow 1$ since it corresponds to certain \textquotedblleft unhealthy
reduction\textquotedblright\ \cite{Horava11}. But some measurements on
quantum Hall effects exhibit the isotropic $z=1$ scaling indeed \cite{Tsui93}%
. Thus, the holographic applications of Ho\v{r}ava gravity are limited. In
this letter we offer an alternative: the bulk is still relativistic without
a preferred time foliation. The global time is only realized at the boundary
by imposing the hypersurface orthogonality condition for vielbein there. The
holographic dictionary ensures the dual field theories are NR, while the
relativistic bulk allows black hole solutions previously well studied in
relativistic holography. This facilitates holographic study of thermal
effects and phase transitions of NR systems. As an example, we employ dyonic
black brane model of \cite{Hartnoll07} to study NR Hall effects with finite
temperature and magnetic field.

\emph{Notations}: We use three types of spacetime indices $M,N,\cdots $, $%
\mu ,\nu ,\cdots $, $i,j,\cdots $, for (3+1)-, (2+1)- and 2-dimensional
manifolds, respectively. We denote the tangent space index by $a,b,\cdots $
for the zweibein. Holographic radial direction is referred to $r$ with
boundary located at $r=0$. \textquotedblleft $\hat{\phantom{a}}$%
\textquotedblright\ and \textquotedblleft $\bar{\phantom{a}}$%
\textquotedblright\ mark the bulk quantities and their boundary values after
stripping off the asymptotic $r$-dependence.

\textbf{\emph{NR Symmetries}}\emph{\ }We consider a (2+1)-dimensional NR
field theory described by the following microscopic action in curves
spacetime
\begin{equation*}
S=\int \!d^{3}x\sqrt{g}\frac{1}{2}\left( i\psi ^{+}\overset{\leftrightarrow }%
{D}_{t}\psi -\frac{g^{ij}+i\varepsilon ^{ij}}{me^{\Phi }}D_{i}\psi ^{\dagger
}D_{j}\psi +\ldots \right)
\end{equation*}%
where \textquotedblleft $\ldots $\textquotedblright\ denotes the
interactions, $g_{ij}$ is spatial metric, $g=\det (g_{ij})$ and
\begin{align}
D_{i}& =\partial _{i}-i\left( A_{i}-s_{0}\omega _{i}\right) , \\
D_{t}& =\partial _{t}-i\left[ A_{t}-s_{0}\omega _{t}+\frac{1}{4me^{\Phi }}%
\left( \mathrm{g}-2\right) B\right] .  \notag
\end{align}%
The magnetic field is $B=\varepsilon ^{ij}\partial _{i}A_{j}$, $\varepsilon
^{ij}=\epsilon ^{ij}/\sqrt{g}$ with $\epsilon ^{ij}$ the Levi-Civita symbol.
$\mathrm{g}$ is the gyromagnetic factor. Our choice of $D_{t}$ such that $(%
\mathrm{g}-2)B/4m$ appears together with $A_{t}$ and $g^{ij}+i\varepsilon
^{ij}$ appears in a combination in the standard Pauli form has been employed
in \cite{Alicki94}. The scalar field $\Phi $ is introduced to source the
energy density \cite{Karch12,Geracie:2014}. The field $\psi $ represents an
underlying microscopic degree of freedom which is to be path-integrated out
when computing the effective action. It can be the electrons as well as
other composite particles. It has intrinsic spin $s_{0}$ and couples to
curved space through spin connection: $\omega _{t}=\frac{1}{2}\epsilon
_{ab}e^{aj}\partial _{t}e_{j}^{b}$, $\omega _{i}=\frac{1}{2}(\epsilon
_{ab}e^{aj}\partial _{i}e_{j}^{b}-\varepsilon ^{jk}\partial _{j}g_{ki})$,
where $e_{i}^{a}$ is the zweibein for metric $g_{ij}$. Note that the spin
connection has been considered recently in \cite{Cho:2014vfl} from flux
attachment. The action is invariant under NR diffeomorphism and Weyl
transformations (parameterized by $\xi ^{\mu }$ and $\sigma $, with $%
\partial _{i}\xi ^{t}=0$) as shown in \cite{Wu1}, with a slightly different
transformation rule for $A_{t}$:%
\begin{align}
\delta A_{t}& =\xi ^{\mu }\partial _{\mu }A_{t}+A_{\mu }\partial _{t}\xi
^{\mu }-\frac{1}{2}\left( 1-s_{0}\right) \varepsilon ^{ij}\partial
_{i}(g_{jk}\partial _{t}\xi ^{k}) \\
& -\frac{1}{4me^{\Phi }}\left( \mathrm{g}-2\right) \left[ \varepsilon
^{ij}\partial _{i}\left( me^{\Phi }g_{jk}\partial _{t}\xi ^{k}\right)
+\left( 1-s_{0}\right) \nabla ^{2}\sigma \right] ,  \notag
\end{align}%
because we do not include Ricci scalar in $D_{t}$. The Ward identities
resulting from the NR spacetime symmetries derived in \cite{Geracie:2014}
are still applicable here.

\textbf{\emph{Global Time}} For the purpose toward NR EFTs, it is convenient
to discuss geometry in term of vielbein rather than metric. For
(2+1)-dimensional relativistic gravity \cite{rel}, $ds^{2}=-\tau ^{2}+\delta
_{ab}e^{a}e^{b}$. The time-like vielbein is $\tau =e^{-\Phi
}(dt-C_{i}dx^{i}) $ and space-like ones $e^{a}=e_{i}^{a}(dx^{i}+N^{i}dt)$.
Notice $g_{ti}=e^{-2\Phi }C_{i}+N_{i}$ where $i$ is lowered by $g_{ij}$. $%
C_{i}$ is the source to NR energy flux \cite{Son08,Karch12,Geracie:2014},
which can be seen by matching its diffeomorphism with that of the source.
However, $C_{i}$ does not appear in the above NR field theory because it is
written in global time coordinates (GTC) where $C_{i}=0$. The condition for
existence of a global time as required by NR causality is $\tau \wedge d\tau
=0$ \cite{Geracie:2014}, which corresponds to the twistless torsion
condition for NC geometry in \cite{Christensen13}. For the application of
relativistic gravity on NR EFTs, this hypersurface orthogonality condition
for vielbein must be imposed. Then we can always work in GTC with $C_{i}=0$.
Under diffeomorphism, $\delta C_{i}=\xi ^{\mu }\partial _{\mu
}C_{i}+C_{j}(\partial _{i}+C_{i}\partial _{t})\xi ^{j}-(\partial
_{i}+C_{i}\partial _{t})\xi ^{t}$. This implies the allowed diffeomorphism
in GTC must satisfies $\partial _{i}\xi ^{t}=0$. This is called foliation
preserving diffeomorphism and is exactly the assumption made to ensure the
diffeomorphism invariance of the above NR field theory. Now $\Phi $ and $%
e_{i}^{a}$ can be identified with their counterparts in NR field theory
because they have the same symmetry transformations. To compute energy flux,
$C_{i}$ dependence has to be restored. This can be done by performing a $%
\partial _{i}\xi ^{t}\neq 0$ diffeomorphism and going away from GTC. For the
rest of this letter, we will work in GTC for simplicity.

\textbf{\emph{Holography and Covariant Map}} For holography in 3+1
dimensions, the bulk theory includes relativistic graviton described by
vielbein $(\hat{\tau},\hat{e}^{a},\hat{n})$ and a $U(1)$ gauge field $\hat{V}%
=\hat{V}_{M}dx^{M}$, among others. We assume the background near boundary $%
r\Rightarrow 0$ is asymptotic Lifshitz with AdS radius $L$:
\begin{equation}
ds^{2}\Rightarrow -\left( L/r\right) ^{2z}dt^{2}+\left( L/r\right)
^{2}\left( d\vec{x}^{2}+dr^{2}\right) ,
\end{equation}%
and choose gauge condition for radial vielbein $\hat{n}=(L/r)dr$ and $\hat{%
\tau}_{r}=\hat{e}_{r}^{a}=V_{r}=0$. These conditions do not completely fix
the bulk gauge freedom. The residual diffeomorphism near boundary is $\hat{%
\xi}^{\mu }\Rightarrow \bar{\xi}^{\mu }$, $\hat{\xi}^{r}\Rightarrow -r\bar{%
\sigma}$. Then near boundary $\hat{\tau}\Rightarrow (L/r)^{z}\bar{\tau}$, $%
\hat{e}^{a}\Rightarrow (L/r)\bar{e}^{a}$ and $\hat{V}_{\mu }\Rightarrow \bar{%
V}_{\mu }$. Under $(\bar{\xi}^{\mu },\bar{\sigma})$, $\bar{\tau}$, $\bar{e}%
^{a}$ and $\bar{V}_{\mu }$ transform in the same way as their counterparts
in (2+1)-dimensional relativistic gravity, hence are identified with above $%
\tau $, $e^{a}$ and a $V_{\mu }$. Now the global time condition becomes a
boundary condition in holography: $\bar{\tau}\wedge d\bar{\tau}=0$. We will
work in the stronger condition $\bar{\tau}_{i}=0$, which forces $\partial
_{i}\bar{\xi}^{t}=0$ at boundary but allows $\partial _{i}\hat{\xi}^{t}\neq
0 $ in the bulk. The map for the $U(1)$ field $\bar{V}_{\mu }=V_{\mu }$ is
non-trivial:%
\begin{align}
V_{i}& =A_{i}-me^{\Phi }N_{i}+s^{\prime }\omega _{i}+\frac{1-s_{0}-s^{\prime
}}{2}\varepsilon ^{jk}g_{ki}\partial _{j}\log (me^{\Phi }),  \notag \\
V_{t}& =A_{t}+\frac{\mathrm{g}-2}{4me^{\Phi }}B-\frac{1}{2}me^{\Phi
}N_{i}N^{i}+s^{\prime }\omega _{t}  \notag \\
& +\frac{1-s_{0}-s^{\prime }}{2}\left[ \varepsilon ^{ij}\partial
_{i}N_{j}+\varepsilon ^{ij}N_{j}\partial _{i}\log (me^{\Phi })\right] ,
\label{general map}
\end{align}%
where $s^{\prime }$ is an arbitrary constant. The additional structures in
the map $\bar{V}_{\mu }=A_{\mu }+\ldots $ are built up to covariantize the
NR gauge field, that is, to cancel the extra terms in $\delta A_{\mu }$ so
that the symmetry transformations on both sides of the map are matched, see
the detail properties of these structures in \cite{Wu1}. Being part of the
holographic dictionary, this is an extension of that in \cite{Karch12} to
parity violating case. It is also the covariant map for (2+1)-dimensional
EFTs (independent of holography), an extension of \cite%
{Son13,Geracie:2014,Andreev:2014gia}. A nice feature is for FQH effect, it
has a good LLL projection when $\mathrm{g}=2$ and $m\rightarrow 0$ (see how
to give the usual LLL constraint on wave function in \cite{Geracie:2014}).
In holography the mass $m$ is dual to a bulk scalar whose near-boundary
behavior matches its Weyl transformation in NR field theory.

\textbf{\emph{Shift Vector}} The only remaining problem in our formalism is
the shift vector $N^{i}$, which has not been interpreted nor determined from
NR field theory point of view. According to its symmetry transformations, it
corresponds to the velocity field in NC formalism \cite{Son13,Geracie:2014}.
In relativistic theories, it sources the momentum density. However, in the
NR theories momentum density $p^{i}$ is completely determined in terms of
charge current $J^{\mu }$ provided that Galilean symmetry is respected and
the particles have the same charge to mass ratio \cite{Greiter89}. From the
Ward identity in flat spacetime, one can read
\begin{equation}
p^{i}=mJ^{i}-\frac{\mathrm{g}-2s_{0}}{4}\epsilon ^{ij}\partial _{j}J^{t}.
\label{eq:momentum-current}
\end{equation}%
Thus $N_{i}$ does not source $p^{i}$ \cite{velocity}. To determine it in
terms of other fields, a constraint has to be imposed. There is no universal
prescription in the literature. \cite{Wu1} has a detailed discussion on how
to impose a diffeomorphism invariant constraint with smooth LLL limit for
FQH effect. Similarly here, there are two viable choices:
\begin{equation}
N_{i}\frac{\delta W}{\delta V_{t}}+g_{ij}\frac{\delta W}{\delta V_{j}}%
=0\quad \text{or}\quad V_{ti}+V_{ij}N^{j}=0,  \label{constrain}
\end{equation}%
where $V_{\mu \nu }=\partial _{\mu }V_{\nu }-\partial _{\nu }V_{\mu }$. The
former is essentially a path-integral in the spirit of \cite{Son13} to
integrate out the velocity field, while the latter also appears recently in
\cite{Andreev:2014gia}. For FQH effect, they yield the same universal
features in Chern-Simons EFT \cite{Wu1}.

\textbf{\emph{Effective Action for FQH states}} As an example, we show how
to build low energy effective action for FQH states using our formalism. We
work in the limit when the magnetic field $B$ is large compared to the
electric field $E_{i}$ and derivatives. At leading order in derivative
expansion, the gauge Chern-Simons term encodes the universal properties. To
build a NR GCI action, we start with a relativistic Chern-Simons term $S_{%
\mathrm{CS}}=\frac{\nu }{4\pi }\int d^{3}x\,V\wedge dV$ and apply the map (%
\ref{general map}), then plug into the constraint (\ref{constrain}) and
solve for $N^{i}$. We get $N^{i}=-\epsilon ^{ij}E_{j}/B+O(\partial
B,\partial \Phi )$. For constant $B$ and $E_{i}$, the final NR GCI
Chern-Simons action has a simple form:
\begin{align}
S_{\mathrm{CS}}& =\frac{\nu }{4\pi }\int d^{3}x\Big\{\left( A+s^{\prime
}\omega \right) \wedge d\left( A+s^{\prime }\omega \right)  \\
& \qquad +\sqrt{g}\left[ \frac{mE^{2}}{B}+\frac{\mathrm{g}-2}{2m}%
B^{2}+O\left( \partial _{\mu }\right) \right] \Big\}  \notag
\end{align}%
where we have set $\Phi =0$ and $O\left( \partial _{\mu }\right) $ denotes
derivative corrections to the local Lagrangian. These corrections can be
systematically calculated to any higher order using (\ref{general map}) and (%
\ref{constrain}), and the first few are given in \cite{Wu1}. All correlators
obtained from this action (for their explicit expressions see also \cite{Wu1}%
), including the derivative corrections omitted in the above expression,
agree with those computed in \cite{Son13}. From this action, we can easily
recognize $s^{\prime }$ as half of the Wen-Zee shift \cite{Wen:1992ej} and
Hall viscosity $\eta _{H}=(\nu B/4\pi )s^{\prime }$. Furthermore, the total
angular momentum density can be extracted from $\int d^{2}\vec{x}\epsilon
_{ij}r^{i}p^{j}$ using (\ref{eq:momentum-current}), after integrating by
parts:%
\begin{equation}
l_{\mathrm{tot}}=-\frac{\nu B}{2\pi }(1-s_{0}).
\end{equation}%
There is another kind of angular momentum density related to the conjugate
of the vorticity in grand potential density \cite{Jensen11,Hoyos1407}. By
identifying $-N^{i}$ with the drift velocity, one can derive the so called
guiding center angular momentum density \cite{Wu1}:%
\begin{equation}
l_{\mathrm{gc}}=-\frac{\nu B}{2\pi }(1-s_{0}-s^{\prime }).
\end{equation}%
Subtracting the latter from the former, we obtain the Landau orbital angular
momentum density $\ell _{\mathrm{orb}}=-(\nu B/2\pi )s^{\prime }$. This
justifies the relation $\eta _{H}=-\ell _{\mathrm{orb}}/2$ in \cite%
{Read:09,Read:11,Nicolis:2011ey}.

There are two more topological terms in (2+1)-dimensional relativistic
theories beside $V\wedge dV$. One is the gravitational Chern-Simons term $%
\mathrm{tr}\left(\tilde{\omega}\wedge d\tilde{\omega}+\frac{2}{3}\tilde{%
\omega}\wedge\tilde{\omega}\wedge\tilde{\omega}\right)$, where $\tilde{\omega%
}$ denotes the non-Abelian spin connection constructed from the full
spacetime vielbein. Its contribution to the NR effective action after
applying the covariant map has been calculated in \cite{Wu1}. Its primary
role is to shift the coefficient of $\omega\wedge d\omega$ by a constant,
which is related to the central charge of chiral conformal field theory on
the boundary and the thermal Hall conductivity. A third relativistic
topological term which mixes the $U(1)$ gauge field with spacetime curvature
had recently been constructed in \cite{Son1403}. By applying our covariant
map, its contributes to NR effective action is equivalent to a shift of the
constant $s^{\prime}$. Local terms in the relativistic parent theory will
generate non-universal features related to the interactions of the
microscopic theory. We will not discuss these terms here.

\textbf{\emph{A Holographic Model}} We now give another example in term of
relativistic holography \cite{Hartnoll07}, which is dual to strongly coupled
quantum fluids in external magnetic fields. The bulk action includes
Einstein-Maxwell terms
\begin{equation*}
\hat{S}_{\mathrm{EM}}=\frac{-2}{\kappa _{4}^{2}}\int d^{4}x\sqrt{-\hat{G}}%
\left( \frac{1}{4}\hat{R}+\frac{3}{2L^{2}}-\frac{L^{2}}{4}\hat{F}_{MN}\hat{F}%
^{MN}\right)
\end{equation*}%
with a non-dynamical Chern-Simons term $\hat{S}_{\mathrm{CS}}=\frac{\nu }{%
16\pi }\int \epsilon ^{MNPQ}\hat{F}_{MN}\hat{F}_{PQ}$. Here $\hat{F}%
_{MN}=\partial _{M}\hat{V}_{N}-\partial _{N}\hat{V}_{M}$. The background
metric is a dyonic black brane
\begin{equation}
\frac{1}{L^{2}}ds^{2}=\frac{\alpha ^{2}}{r^{2}}\left[ -f\left( r\right)
dt^{2}+d\vec{x}^{2}\right] +\frac{dr^{2}}{r^{2}f(r)},
\end{equation}%
with $\hat{V}_{t}=-q\alpha (1-r)$, $\hat{V}_{y}=h\alpha ^{2}x$, where $q$
and $h$ are electric and magnetic charges. $h$ is related to a constant
magnetic field $B_{0}=h\alpha ^{2}$ on the boundary. The blackening function
is $f\left( r\right) =1-(1+h^{2}+q^{2})r^{3}+\left( h^{2}+q^{2}\right) r^{4}$%
, with the horizon located at $r=1$. The mass parameter $\alpha $ is related
to the Hawking temperature by $4\pi T=\alpha \left( 3-h^{2}-q^{2}\right) .$
The renormalized action is given by subtracting the Gibbons-Hawking term and
a counter-term of the boundary volume. To calculate correlation functions,
we solve all the nine metric and gauge fluctuations in the bulk up to linear
order in momentum $(\omega ,\vec{k})$. By rotational symmetry we can set $%
k_{x}=k$, $k_{y}=0$. We will not list the full expressions of the solutions
nor the action here, but only give results of correlators computed from
them. The procedure is similar to that in \cite{Hartnoll07}. After obtaining
the on-shell boundary action, still in relativistic form, we solve the
constraint equation (\ref{constrain}) (we use the first one) and apply the
holographic dictionary (\ref{general map}) to calculate the NR GCI effective
action.

The non-vanishing 1-point functions are the charge density $\rho $, energy
density $\epsilon $ and internal pressure $P$:
\begin{align}
\rho & =\frac{\nu B_{0}}{2\pi }-\frac{2L^{2}}{\kappa _{4}^{2}}q\alpha ^{2},
\\
\epsilon & =-\frac{\left( \mathrm{g}-2\right) \rho B_{0}}{4m}+\frac{L^{2}}{%
\kappa _{4}^{2}}(1+h^{2}+q^{2})\alpha ^{3},  \notag \\
P& =-\frac{\left( \mathrm{g}-2\right) \rho B_{0}}{4m}+\frac{1}{2}\frac{L^{2}%
}{\kappa _{4}^{2}}(1+h^{2}+q^{2})\alpha ^{3}.  \notag
\end{align}%
Thermodynamic pressure (i.e. the grand potential density) can be obtained
from the background action:
\begin{equation}
P_{\mathrm{thm}}=\frac{L^{2}}{2\kappa _{4}^{2}}(1-3h^{2}+q^{2})\alpha ^{3}-%
\frac{\nu B_{0}}{2\pi }q\alpha .
\end{equation}%
The chemical potential $\mu =-q\alpha -\left( \mathrm{g}-2\right) B_{0}/(4m)$
is identified with the background of $A_{t}$ in the NR field theory. The
magnetization density is defined as $M=\left. \partial P_{\mathrm{thm}%
}/\partial B_{0}\right\vert _{T,\mu }$. Bekenstein-Hawking law gives entropy
density $s=2\pi L^{2}\alpha ^{2}/\kappa _{4}^{2}$. All these thermodynamic
quantities satisfy the following fundamental relation:
\begin{equation}
\epsilon +P-Ts-\mu \rho +B_{0}M=0.
\end{equation}%
The system has local thermodynamic stability:
\begin{equation*}
\det \left[ \partial _{\rho }\partial _{s}\epsilon \left( \rho ,s\right) %
\right] =\frac{12s^{2}+48\pi ^{2}B_{0}^{2}+\left( 2\pi \rho -\nu
B_{0}\right) ^{2}}{64\pi s^{2}}>0.
\end{equation*}

Some of the 2-point functions are%
\begin{align}
G_{\mathrm{ra}}^{1,1}& =G_{\mathrm{ra}}^{2,2}=0,\quad G_{\mathrm{ra}%
}^{1,2}=-G_{\mathrm{ra}}^{2,1}=\frac{i\omega \rho }{B_{0}}, \\
G_{\mathrm{ra}}^{1,\Phi }& =0,\quad G_{\mathrm{ra}}^{2,\Phi }=-ik\frac{%
\epsilon +P}{B_{0}},  \notag \\
G_{\mathrm{ra}}^{11,2}& =\frac{-ikq\alpha ^{2}\left( \mathrm{g}-2\right) }{2m%
}+\frac{L^{2}}{\kappa _{4}^{2}}\frac{3ik}{8B_{0}}(1+h^{2}+q^{2})\alpha ^{3},
\notag
\end{align}%
where definitions of retarded correlators $G_{\mathrm{ra}}^{A,B}$ follow
\cite{Geracie:2014}. From $G_{\mathrm{ra}}^{i,j}$, the longitudinal
conductivity vanishes and Hall conductivity equals $\rho /B_{0}$ as
expected. $G_{\mathrm{ra}}^{i,\Phi }$ shows current response to
inhomogeneous gravitational field $\partial _{i}\Phi $, with a transport
coefficient $\sigma _{H}^{G}=\left( \epsilon +P\right) /B_{0}$ that agrees
with \cite{Geracie:2014 2} from hydrodynamic analysis on LLL. The above
correlators satisfy Ward identities given in Eqs. (47) and (48) in \cite%
{Geracie:2014}.

The shear, bulk and Hall viscosities are
\begin{equation}
\eta =\frac{s}{4\pi },\quad \zeta =\frac{\kappa _{\mathrm{int}}^{-1}}{%
i\omega },\quad \eta _{H}=\frac{1}{2}\rho s^{\prime }.
\end{equation}%
Some remarks are in order. First, the shear viscosity characterizes the
fluid is strongly coupling. Second, \cite{Bradlyn:2012,Geracie:2014} pointed
out that there is a zero-frequency divergent term proportional to the
inverse internal compressibility $\kappa _{\mathrm{int}}^{-1}$ in the bulk
viscosity. For our case, it is
\begin{equation}
\kappa _{\mathrm{int}}^{-1}=\frac{3}{4}(1+h^{2}+q^{2})\alpha ^{3}-\frac{%
\left( \mathrm{g}-2\right) \rho B_{0}}{2m}.
\end{equation}%
Other than this contact term, the bulk viscosity is vanishing as required by
Weyl invariance. Third, the covariant map (\ref{general map}) is crucial to
the above non-vanishing Hall viscosity, whose form in terms of charge
density and the shift agrees with \cite{Read:09,Read:11}. Moreover, by
calculating the total angular momentum density
\begin{equation}
l_{\mathrm{tot}}=\frac{\left( \mathrm{g}-2\right) }{8\pi }\left( \nu
B_{0}+4\pi \rho \right) -\frac{\mathrm{g}-2s_{0}}{2}\rho
\end{equation}%
and the guiding center one%
\begin{equation}
l_{\mathrm{gc}}=\frac{\left( \mathrm{g}-2\right) }{8\pi }\nu
B_{0}-(1-s_{0}-s^{\prime })\rho ,
\end{equation}%
we can check $\ell _{\mathrm{orb}}=-\rho s^{\prime }$, which indicates $\eta
_{H}=-\ell _{\mathrm{orb}}/2$ in the strongly coupling model.

At the end we have two comments on the mass $m$. (1) The dyonic black brane
background is asymptotic AdS with dynamical exponent $z=1$. From our
holographic dictionary, for $z\neq2$, $m$ is dual to a bulk scalar with
non-trivial profile and $r^{z-2}$ asymptote. Here we do not consider this
profile explicitly because we work in the probe limit where this scalar
sector can be engineered such that it decouples, similar as in \cite{Karch12}%
. (2) Instead of working in the probe limit, we can also project to LLL,
where $m\rightarrow0$ and $\mathrm{g}=2$. In this limit the cyclotron
frequency $\omega_{c}=B/m$ diverges which forbids higher Landau level
mixing. In this case all $(\mathrm{g}-2)/m$ terms in the above expressions
drop off, and this holographic model becomes one for LLL.

\textbf{\textit{Conclusions}} We have shown that relativistic gravity
theories can be used as a framework to build effective theories for NR
systems that respect all NR spacetime symmetries, holographically or not,
for any dynamical exponent $z$. In order to adapt to the global time, the
time-like vielbein must satisfy hypersurface orthogonality condition. Under
this condition, we present a covariant map that relates the relativistic
gauge field to the NR one, which can also serve as part of the holographic
dictionary. Additional constraints are given to fix the shift vector. Our
formalism is particular suitable for spin-polarized NR particles, including
the FQH fluids and chiral superfluids \cite{Moroz1305}. Low energy effective
actions for these systems are then constructed from purely (2+1)-dimensional
Chern-Simons field theory and from (3+1)-dimensional holographic theory with
a dyonic black brane background. They have a good LLL projection and capture
the linear response properties of these systems.

\textbf{\textit{Acknowledgments}} We are very grateful to Dam Thanh Son for
constant support and inspiring discussions throughout the progress of this
work. We thank Andreas Karch for reading the draft and Michael Geracie for
useful discussions. C. Wu is supported by the US DOE grant No.
DE-FG02-13ER41958 and S.-F. Wu was supported by NNSFC No. 11275120 and the
China Scholarship Council.


\begin{thebibliography}{99}
\bibitem{rel} In non-holographic context, the \textquotedblleft relativistic
gravity\textquotedblright\ means the relativistic background geometry while
not involving the spacetime dynamics.
\bibitem{velocity} Alternatively, one would introduce the shift vector as
the source of momentum density in the microscopic action. In a Galilean
invariant theory, it will appear only in a specific combination with the
source of number current [13]. Consequently, the form of the effective
action will be constrained. For instance, the holographic model that we will
consider below is not applicable to provide the desired action.
\end{thebibliography}
\end{document}